\documentclass[conference,letterpaper]{IEEEtran}

\usepackage{siunitx}
\usepackage{booktabs}
\usepackage{comment}

\usepackage[utf8]{inputenc} 
\usepackage[T1]{fontenc}
\usepackage{url}
\usepackage{ifthen}
\usepackage{cite}
\usepackage[cmex10]{amsmath}   
\usepackage{cases}
\usepackage{multirow}
\usepackage{empheq}
\usepackage{verbatim}
\usepackage{algorithm}
\usepackage {algpseudocode}
\usepackage{algorithmicx}
\usepackage{algcompatible}
\usepackage{amssymb}
\usepackage{optidef}
\ifCLASSINFOpdf
  \usepackage[pdftex]{graphicx}
\else
  \usepackage[dvips]{graphicx}
\fi
\usepackage[caption=false,font=footnotesize]{subfig}

\usepackage{color}
\usepackage{multirow}

\newtheorem{lemma}{Lemma}

\newtheorem{proposition}{Proposition}

\begin{document}
\title{Lowering the Error Floor of Error Correction Code Transformer} 

\author{%
    \IEEEauthorblockN{Taewoo Park\textsuperscript{*}, Seong-Joon Park\textsuperscript{\textdagger}, Hee-Youl Kwak\textsuperscript{\textdaggerdbl}, Sang-Hyo Kim\textsuperscript{\S}, and Yongjune Kim\textsuperscript{*}}
    \IEEEauthorblockA{\textsuperscript{*}Department of Electrical Engineering and \textsuperscript{\textdagger}Institute of Artificial Intelligence, POSTECH, Pohang 37673, South Korea \\
    Email: \{parktaewoo, seongjoon, yongjune\}@postech.ac.kr\\
    \textsuperscript{\textdaggerdbl}Department of Electrical, Electronic and Computer Engineering, University of Ulsan, Ulsan 44610, South Korea\\
    Email: hykwak@ulsan.ac.kr\\
    \textsuperscript{\S}Department of Electrical and Computer Engineering, Sungkyunkwan University, Suwon 16419, South Korea\\
    Email: iamshkim@skku.edu
    }
}

\maketitle

\begin{abstract}
    With the success of transformer architectures across diverse applications, the error correction code transformer (ECCT) has gained significant attention for its superior decoding performance. 
    In spite of its advantages, the error floor problem in ECCT decoding remains unexplored.
    We present the first investigation into this issue, revealing that ECCT encounters error floors, limiting its effectiveness in practical settings. 
    To address this error floor problem, we adopt a hybrid decoding framework that integrates ECCT with conventional hard decision decoders. 
    Unlike prior hybrid decoding schemes, our key contribution lies in proposing a novel loss function that explicitly takes into account the interaction between ECCT and hard decision decoders during training. 
    The proposed loss function guides ECCT to focus on residual errors that are not corrected by the hard decision stages, effectively lowering the error floor. 
    Simulation results confirm that the hybrid decoder trained with the proposed loss function achieves substantial performance gains over standard ECCT in both the waterfall and the error floor regions.

\end{abstract}

\section{Introduction}
    


    The transformer architecture~\cite{Vaswani2017attention} has achieved remarkable success across a range of applications~\cite{Dosovitskiy2021image, Devlin2018bert}, motivating the development of a transformer-based error correction code decoder, known as the error correction code transformer (ECCT)~\cite{Choukroun2022error}. 
    ECCT has achieved notable decoding performance, particularly for short codes, by leveraging the attention mechanism to capture inter-bit dependencies.
    Building on this approach, Choukroun and Wolf proposed a foundation model for ECCT~\cite{Choukroun2024foundation} and an end-to-end optimization framework to jointly train encoder and decoder~\cite{Choukroun2024learning}.
    Park~\textit{et al.} introduced novel architectures based on multiple mask matrices~\cite{Park2025multiple} and the cross-attention mechanism~\cite{Park2025crossmpt}, achieving state-of-the-art decoding performance among neural decoders.

    
    In spite of its notable decoding performance in the waterfall region, the error floor of ECCT has yet to be investigated. 
    The error floor refers to the phenomenon where the decrease in error probability slows significantly in the high signal-to-noise ratio (SNR) region, compared to the waterfall region~\cite{Richardson2003error}.    
    For emerging applications demanding extremely low error rate, such as the next-generation ultra-reliable and low-latency communications~(xURLLC)~\cite{Hong20226g}, it is important to mitigate the error floor problem~\cite{Kwak2023boosting}.
    In particular, several techniques have been developed to mitigate the error floor phenomenon in model-based decoders for low-density parity-check~(LDPC) codes by leveraging decoder diversity~\cite{Xiao2021faid} and boosting-based learning~\cite{Kwak2023boosting,Kwak2025boosted}.
    However, to the best of our knowledge, the error floor in transformer-based decoders remains unexplored.

    In this paper, we conduct the first investigation of the error floor phenomenon in transformer-based decoders, which has not been previously reported or analyzed. 
    We discover that ECCT suffers from error floors in both Bose-Chaudhuri-Hocquenghem~(BCH) and LDPC codes, limiting its effectiveness at achieving low frame error rates (FERs) in the high SNR regime.
    Notably, ECCT exhibits an error floor for BCH codes, despite the fact that conventional decoders do not encounter this problem~\cite{Ryan2009channel}. 
    For LDPC codes, we observe that ECCT inherits the well-known error floor behavior of the belief propagation (BP) algorithm.
    These findings motivate the development of decoding strategies to address the error floor in ECCT. 
    
    To address the error floor in ECCT, we adopt a simple yet effective hybrid decoding strategy that integrates ECCT with hard decision decoders as pre- and post-decoders.
    Although hybrid decoding has been widely explored~\cite{Declercq2013finite, Nguyen2014two, Kang2016breaking, Kwak2023boosting, Xiao2021faid, Kim2025hybrid}, our key contribution lies in the design of a novel loss function that explicitly accounts for the integration between ECCT and hard decision components. 
    This loss function guides the model parameters of ECCT to focus on residual errors not corrected by the hard decision stages, thereby effectively lowering the error floor.
    Specifically, to account for the pre-decoder, the training dataset is restricted to received vectors with error patterns beyond the error correction capability of the hard decision decoder.
    By taking into account the post-decoder, the loss function is adjusted depending on whether the number of errors in ECCT's output exceeds the error correction capability.
    Unlike~\cite{Choukroun2022error,Choukroun2024foundation,Choukroun2024learning,Park2025multiple,Park2025crossmpt}, which train ECCT using the standard binary cross entropy loss, the proposed loss function directs ECCT to prioritize correcting critical errors within the hybrid decoding strategy.
    This targeted training improves the overall effectiveness of the hybrid decoding process.

    \begin{figure*}[!t] 
        \centering
        \subfloat[BCH codes]{\includegraphics[width=0.4\textwidth]{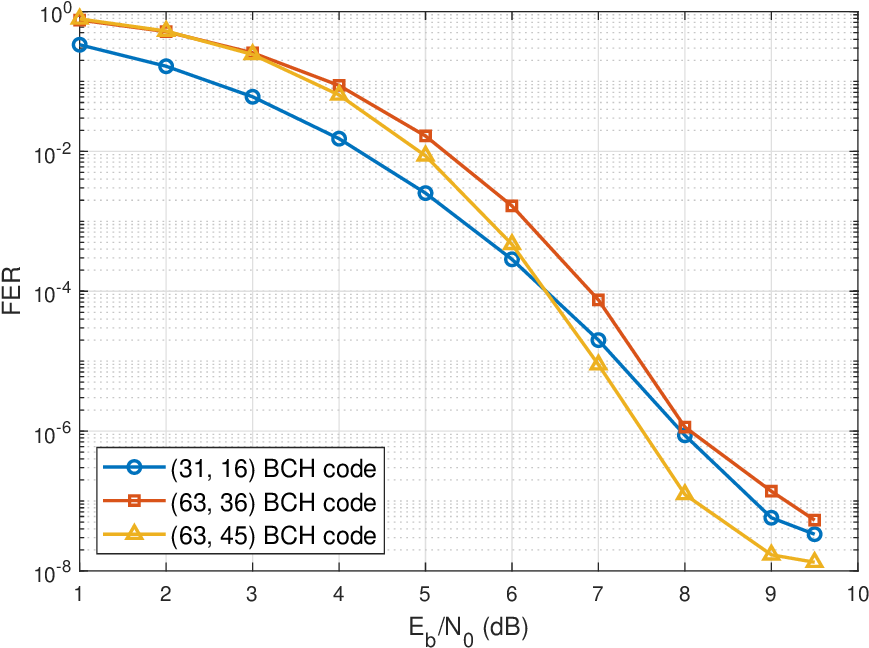}\label{fig:errorfloor_bch}}
        \hfil
        \subfloat[LDPC codes]{\includegraphics[width=0.4\textwidth]{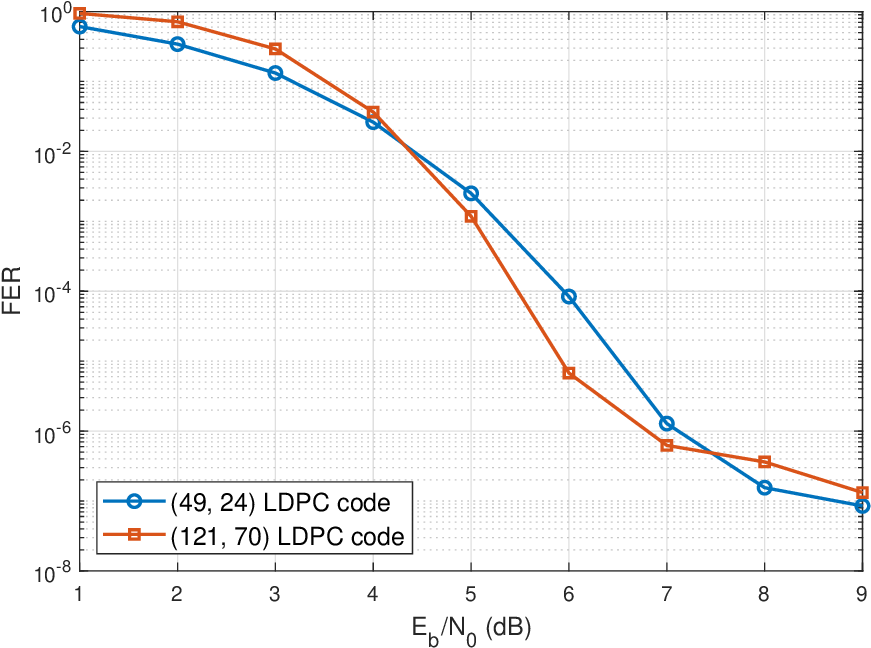}\label{fig:errorfloor_ldpc}}
        \hfil
        \caption{Error floor phenomenon of ECCT: (a) FER performance of ECCT for BCH codes, and (b) FER performance of ECCT for LDPC codes.}
        \label{fig:errorfloor}
        \vspace{-4mm}
    \end{figure*}

    We evaluate the decoding performance of the hybrid decoder trained with the proposed loss function on BCH codes.
    Experimental results demonstrate that the proposed method effectively mitigates the error floor phenomenon and achieves notable performance gains even in the waterfall region.        

    \section{Preliminaries}\label{sec:preliminaries}
    
    \subsection{Notation}
    We denote an $(n,k)$ linear block code as $\mathcal{C}$, which is characterized by a generator matrix $G$, a parity check matrix $H$, and a minimum Hamming distance $d_{\text{min}}$.
    The message and the codeword are denoted by $\mathbf{m}\in\{0,1\}^k$ and $\mathbf{x}\in \mathcal{C} \subseteq \{0,1\}^n$.    
    Additionally, $\mathbf{X}$ represents the random variable of $\mathbf{x}$.
    We consider the binary phase shift keying (BPSK) modulation with the additive white Gaussian noise (AWGN) channel. 
    The modulated signal and the received vector are denoted by $\mathbf{x}_s$ and $\mathbf{y}$ with the noise vector $\mathbf{z}\sim\mathcal{N}(0,\sigma^2I)$.
    The Hamming distance between two codewords $\mathbf{x}_1$ and $\mathbf{x}_2$ is denoted by $d_H(\mathbf{x}_1,\mathbf{x}_2)$.
    The decoder $f(\cdot)$ outputs the estimated codeword $\widehat{\mathbf{x}}$ from the received vector $\mathbf{y}$.
    The hard decision decoder corrects any received vector $\mathbf{y}$ whose hard decision version $\tilde{\mathbf{y}}\in\{0,1\}^n$ satisfies $d_H(\mathbf{x},\tilde{\mathbf{y}})\leq t_c$, where the error correction capability is denoted as $t_c = \lfloor (d_{\text{min}}-1)/2\rfloor$.    

    \subsection{Error Correction Code Transformer}\label{sec:ecct}
    ECCT is a model-free neural decoder based on the transformer architecture~\cite{Choukroun2022error}.
    As in~\cite{Bennatan2018deep}, ECCT preprocesses the input as \mbox{$h(\mathbf{y})=[|\mathbf{y}|,s(\mathbf{y})]$}, where $[\cdot,\cdot]$ denotes the concatenation of two vectors and $s(\mathbf{y})$ represents the syndrome \mbox{$H\tilde{\mathbf{y}}=H\text{bin}(\text{sign}(\mathbf{y}))$}. 
    Here, $\text{sign}(\cdot)$ is the sign function such that $\text{sign}(\alpha)=1$ if $\alpha\geq 0$ and $-1$ otherwise; and $\text{bin}(\cdot)$ is the binarization function defined as $\text{bin}(\alpha)=0.5(1-\alpha)$.
    Each element of the preprocessed input is then linearly mapped into an embedding vector of dimension $d$ for the attention layers.

    The code-aware self-attention operation is defined as~\cite{Choukroun2022error}:
    \begin{equation}
        A_H\left(Q,K,V\right)=\text{Softmax}(d^{-1/2}(QK^{\mathsf{T}}+{g}(H)))V,\nonumber
    \end{equation}
    where $Q$, $K$, $V$, and ${g}(H)$ denote the query, key, value, and mask matrix, respectively.
    The mask matrix $g(H)$ is constructed using the parity check matrix $H$, enabling the self-attention to focus selectively on the important relations between the embedding vectors.

    The ECCT decoder, composed of $N$ decoding layers that consist of code-aware self-attention and feed-forward neural network, is denoted as $f_{\boldsymbol{\theta}}(\mathbf{y})$ with trainable parameters $\boldsymbol{\theta}$, and aims to predict the multiplicative noise $\tilde{\mathbf{z}}_s$~\cite{Choukroun2022error}. 
    The multiplicative noise is defined by  $\mathbf{y}=\mathbf{x}_s+\mathbf{z}=\mathbf{x}_s\odot\tilde{\mathbf{z}}_s$, where $\odot$ denotes element-wise multiplication.
    ECCT is trained using the following binary cross entropy loss function:
    \begin{align}\label{eq:bce}
        l_\text{BCE}(\mathbf x,f_{\boldsymbol{\theta}}(\mathbf y)) &=-\sum_{i=1}^n \{\tilde z_i\log(1-\sigma(f_{\boldsymbol{\theta}}(\mathbf y)_i)) \nonumber \\
        &\quad +(1-\tilde z_i)\log \sigma(f_{\boldsymbol{\theta}}(\mathbf{y})_i)\},
    \end{align}
    where $\tilde{\mathbf{z}}=\text{bin}(\text{sign}(\tilde{\mathbf{z}}_s))$ denotes the binarized multiplicative noise and $\sigma(\cdot)$ denotes the sigmoid function $\sigma(\alpha)=1/(1+e^{-\alpha})$.
    Lastly, the estimated codeword can be obtained as 
    \begin{equation}
        \widetilde{f}_{\boldsymbol{\theta}}(\mathbf{y})=\text{bin}(\text{sign}(f_{\boldsymbol{\theta}}(\mathbf{y})\odot \mathbf{y})).
    \end{equation}

    \section{Error Floor of Error Correction Code Transformer and Hybrid Decoders}
        
    We report the error floor phenomenon of ECCT for the first time in the literature. 
    Fig.~\ref{fig:errorfloor} shows the error floors of ECCT with $N=6$ and $d=128$ when decoding BCH and LDPC codes.
    ECCT suffers from error floors for both BCH and LDPC codes, limiting its ability to achieve extremely low FER in the high SNR region.
    Notably, ECCT exhibits an error floor for BCH codes, even though conventional decoders do not encounter this issue.
    For LDPC codes, ECCT also fails to overcome the error floor, similar to the behavior observed with BP algorithms.
    These findings indicate that transformer-based decoders are susceptible to the error floor phenomenon. 
    
    \begin{figure*}[t] 
        \centering
        \subfloat[Hybrid decoding with hard decision pre- and post-decoders]{\includegraphics[width=0.9\textwidth]{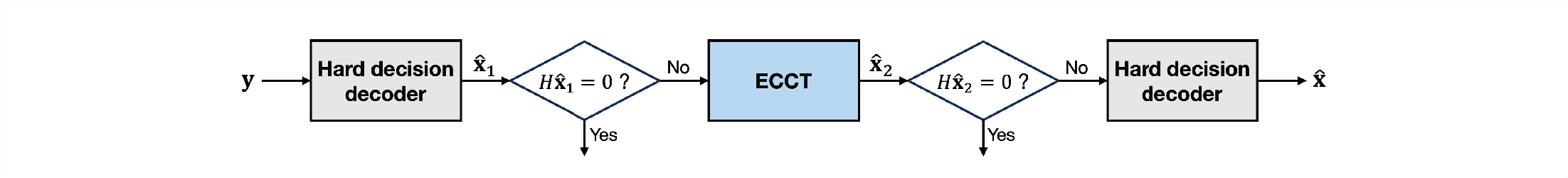}\label{fig:architecture_pre_post}}
        \hfil
        \subfloat[Hybrid decoding with hard decision pre-decoder]{\includegraphics[width=0.45\textwidth]{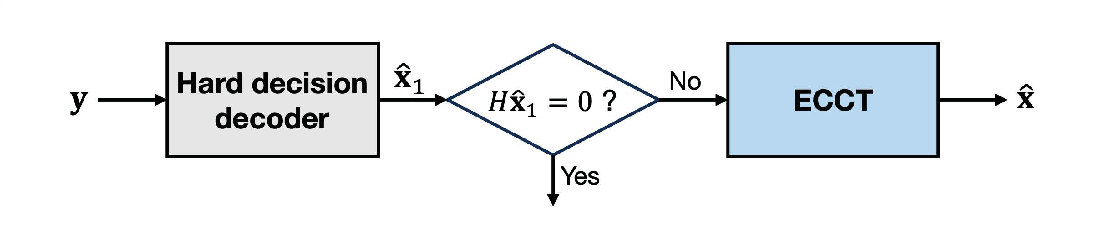}\label{fig:architecture_pre}}
        \hfil
        \subfloat[Hybrid decoding with hard decision post-decoder]{\includegraphics[width=0.45\textwidth]{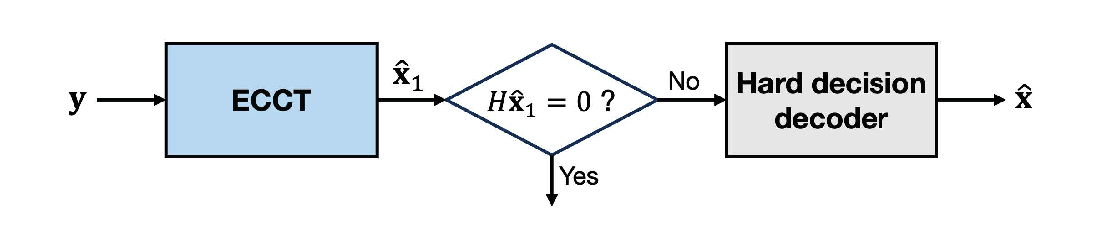}\label{fig:architecture_post}}
        \hfil
        \caption{Block diagrams of hybrid decoding architectures.}
        \label{fig:architecture}
        \vspace{-4mm}
    \end{figure*}
        
    
    We adopt a hybrid decoding strategy that combines ECCT with hard decision decoders to lower the error floor of ECCT.
    The block diagrams of the proposed architectures are shown in Fig.~\ref{fig:architecture}.
    As shown in Figs.~\ref{fig:architecture}\subref{fig:architecture_pre_post}, \subref{fig:architecture_pre}, and \subref{fig:architecture_post}, hard decision decoders can be employed as both pre- and post-decoder, or solely as a pre-decoder or a post-decoder, respectively.
    Since hard decision decoders and ECCT can correct different classes of error patterns, the hybrid decoding approach can correct a broader range of errors, effectively lowering the error floor.
    
    In the hybrid decoding architecture with hard decision pre- and post-decoders (see Fig.~\ref{fig:architecture}\subref{fig:architecture_pre_post}), the decoding process terminates if the hard decision pre-decoder successfully finds a valid codeword; otherwise, it proceeds to the next stage.
    When the syndrome of the estimated codeword is not zero, the decoder declares failure.
    If the pre-decoder fails, the received vector $\mathbf{y}$ is passed to ECCT for decoding in the subsequent stage. 
    If ECCT also fails, the process continues to the hard decision post-decoder.
    Unlike the pre-decoder, which operates on the received vector $\mathbf{y}$, the post-decoder takes the output of ECCT as its input. 
    The ECCT output typically contains fewer errors since ECCT can reduce errors even when it fails to decode.
    Depending on the situation, either the pre-decoder or the post-decoder can be omitted.
    
    \begin{figure}[t] 
        \centering
        \subfloat[$d_H(\mathbf{x},\tilde{\mathbf{y}})\leq t_c$]{\includegraphics[width=0.36\textwidth]{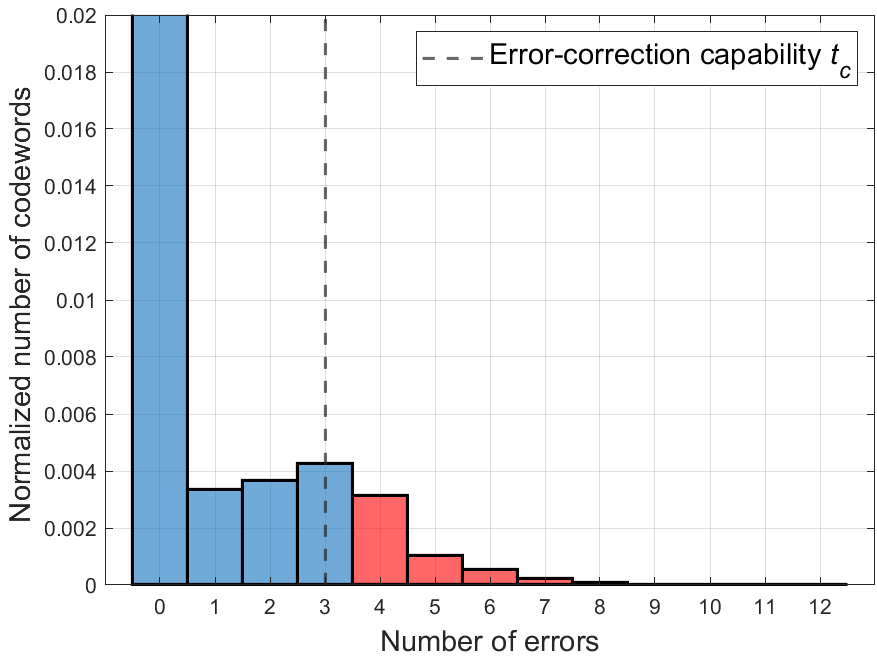}\label{fig:error_distribution-a}}
        \hfil
        \subfloat[$d_H(\mathbf{x},\tilde{\mathbf{y}}) > t_c$]{\includegraphics[width=0.36\textwidth]{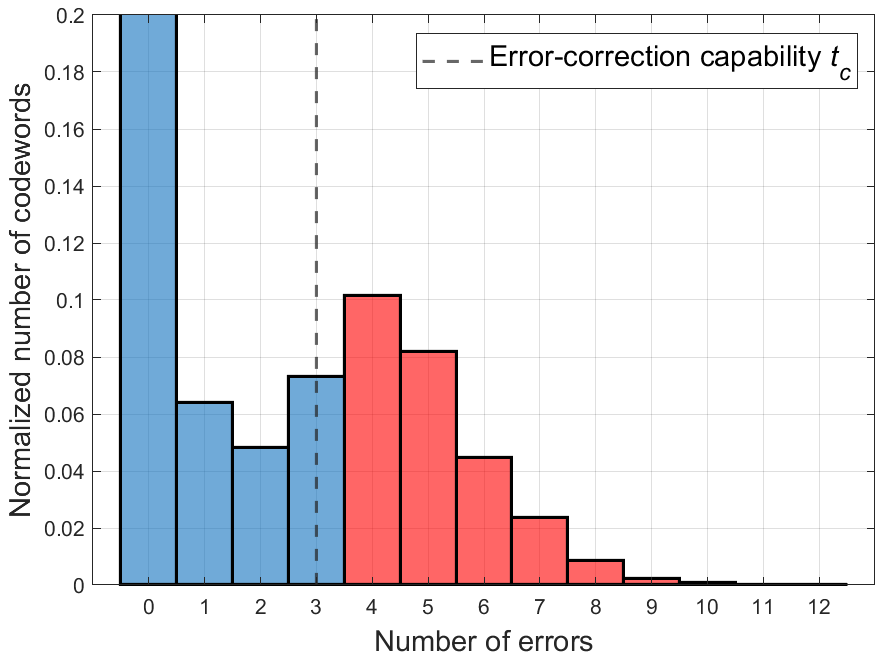}\label{fig:error_distribution-b}}
        \caption{Histograms of the number of errors in the output vector of ECCT in decoding (63, 45) BCH codes. In each figure, blue bars represent cases where the number of errors in ECCT output is less than or equal to $t_c$, while red bars indicate cases where the number of errors exceeds $t_c$.}
        \label{fig:error distribution}
        \vspace{-4mm}
    \end{figure} 
    
    The roles of the pre-decoder and post-decoder in hybrid decoding strategy can be better understood by examining the distribution of errors in the output of ECCT.
    Fig.~\ref{fig:error distribution} shows the histogram of the number of errors in the output vectors of ECCT for the $(63, 45)$ BCH code at \SI{4}{\decibel}.
    Figs.~\ref{fig:error distribution}\subref{fig:error_distribution-a} and \subref{fig:error_distribution-b} show the histogram of received vectors with errors less than or equal to $t_c$ and the histogram of received vectors with errors more than $t_c$, respectively.
    As shown in Fig.~\ref{fig:error distribution}\subref{fig:error_distribution-a}, ECCT may fail to correct all errors, in some cases, even increase the number of errors for received vectors that initially have $t_c$ or fewer errors, that is illustrated with red bars.
    Since the hard decision decoder can correct all the received vectors with $t_c$ or fewer errors, using it as a pre-decoder can improve the overall decoding performance.
    For some received vectors with more than $t_c$ errors, ECCT is capable of estimating the correct codewords. 
    Even when it fails, it frequently reduces the number of errors as shown in Fig.~\ref{fig:error distribution}\subref{fig:error_distribution-b}.
    This behavior occurs since the binary cross entropy loss~\eqref{eq:bce} trains ECCT to reduce bit errors in its output. 
    Consequently, employing a hard decision post-decoder can correct the remaining errors ($\leq t_c$), thereby lowering the error floor.



    \section{Proposed Loss Function for Hybrid Decoders}\label{sec:training}
    
    This section presents a training method for the hybrid decoding strategy.
    To improve the error correction performance of the hybrid decoder, the ECCT decoder should be trained with consideration of its integration with hard decision decoders.
    We first focus on the hybrid decoding scheme that combines ECCT with both pre- and post-decoders, as shown in Fig.~\ref{fig:architecture}\subref{fig:architecture_pre_post}.
    The remaining configurations illustrated in Figs.~\ref{fig:architecture}\subref{fig:architecture_pre} and \subref{fig:architecture_post} are then described, along with a guideline to select the appropriate one for a given scenario.
    Suppose that $\widehat{\mathbf{x}}_1 = f_{\text{pre}}(\mathbf{y})$, $\widehat{\mathbf{x}}_2 = \widetilde{f}_{\boldsymbol{\theta}}(\widehat{\mathbf{x}}_1)$, and $\widehat{\mathbf{x}} = f_{\text{post}}(\widehat{\mathbf{x}}_2)$ represent the outputs of the pre-decoder, ECCT, and the post-decoder, respectively.
    The bit error rate~(BER) loss for this hybrid decoder is given by
    \begin{align}
        \mathcal{L}(\boldsymbol{\theta})&=\mathbb{E}\left[d_H(\mathbf{X},\widehat{\mathbf{X}})\right] \nonumber \\
        &=\mathbb{E} \Big[d_H(\mathbf{X}, f_{\text{post}}\circ f_{\boldsymbol{\theta}}\circ f_{\text{pre}}(\mathbf{Y}) )\Big].\label{eq:loss_hybrid2}
    \end{align}
    
    However, ECCT cannot be trained directly using the loss~\eqref{eq:loss_hybrid2} since the hard decision decoders $f_{\text{pre}}$ and $f_{\text{post}}$ are not differentiable.
    To enable gradient-based training of ECCT, we derive the following equivalent loss for hybrid decoding: 
    \begin{lemma}\label{lemma:loss_hard}
        Consider the hybrid decoding algorithm with the hard decision decoders as pre- and post-decoders. The loss~\eqref{eq:loss_hybrid2} is equivalent to:
        \begin{align}
            P_{\text{dd}}\mathbb{E} \Big[ &
                u(d_H (\mathbf{X}, \widetilde{f}_{\boldsymbol{\theta}}(\mathbf{Y})) - t_c) \nonumber \\
                &   \times d_H (\mathbf{X}, \widetilde{f}_{\boldsymbol{\theta}}(\mathbf{Y}) ) \Big|
                \mathbf{s}_{\text{pre}}\neq \mathbf{0} \Big] + P_{\text{bu}},\label{eq:lemma1}
        \end{align}
        where $\mathbf{s}_{\text{pre}} = Hf_{\text{pre}}(\mathbf Y)$. 
        Also, $P_{\text{bu}}$ and $P_{\text{dd}}$ denote the bit error probability due to undetected errors and the decoding failure probability for detected errors, respectively, for the hard decision decoder. 
        $u(\alpha)$ denotes the step function, defined as $u(\alpha)=0$ if $\alpha \leq 0$ and $1$ otherwise.
    \end{lemma}
    \begin{IEEEproof}
        The expectation in~\eqref{eq:loss_hybrid2} can be evaluated by considering three events: (i) $d_H(\mathbf X,\widetilde{\mathbf Y})\leq t_c$; (ii) $d_H(\mathbf X,\widetilde{\mathbf Y})> t_c$ and $\mathbf{s}_{\text{pre}}\neq \mathbf{0}$; and (iii) $d_H(\mathbf X,\widetilde{\mathbf Y})> t_c$ and $\mathbf{s}_{\text{pre}}= \mathbf{0}$.
        Using the law of total expectation, we can expand the loss as follows:
        \begin{align}
            \mathcal{L}(\boldsymbol{\theta})
            &=\mathbb E\left[d_H(\mathbf X, \widehat{\mathbf{X}}) \Big| d_H(\mathbf X,\widetilde{\mathbf Y}) > t_c, \mathbf{s}_{\text{pre}}\neq \mathbf{0}\right]  \nonumber\\
            & \quad \times \Pr(d_H(\mathbf{X},\widetilde{\mathbf{Y}})>t_c, \mathbf{s}_{\text{pre}} \neq \mathbf{0})\nonumber\\
            & \quad + \mathbb E\left[d_H(\mathbf X,\widehat{\mathbf{X}}) \Big| d_H(\mathbf X,\widetilde{\mathbf Y}) > t_c, \mathbf{s}_{\text{pre}} = \mathbf{0}\right]\nonumber\\
            &\quad \times \Pr(d_H(\mathbf{X},\widetilde{\mathbf{Y}})>t_c, \mathbf{s}_{\text{pre}} = \mathbf{0})\label{eq:total_exp_1}\\
            &=P_{\text{dd}}\mathbb E\Big[d_H(\mathbf X, f_{\text{post}}\circ f_{\boldsymbol{\theta}}(\mathbf Y)) \Big|\mathbf{s}_{\text{pre}}\neq \mathbf{0}\Big] +P_{\text{bu}},\label{eq:total_exp_2}
        \end{align}
        where~\eqref{eq:total_exp_1} holds since the hard decision pre-decoder can correct all errors whenever $d_H(\mathbf{x},\tilde{\mathbf{y}})\leq t_c$, and \eqref{eq:total_exp_2} holds because the pre-decoder passes the received vector to the subsequent stage if it fails.
        Since the decoding success of the post-decoder depends only on the number of errors in the ECCT output, the loss can be expressed as~\eqref{eq:lemma1} using the step function. 
    \end{IEEEproof}

    \begin{figure*}[!t] 
        \centering
        \subfloat[(31, 16) BCH code]{\includegraphics[width=0.4\textwidth]{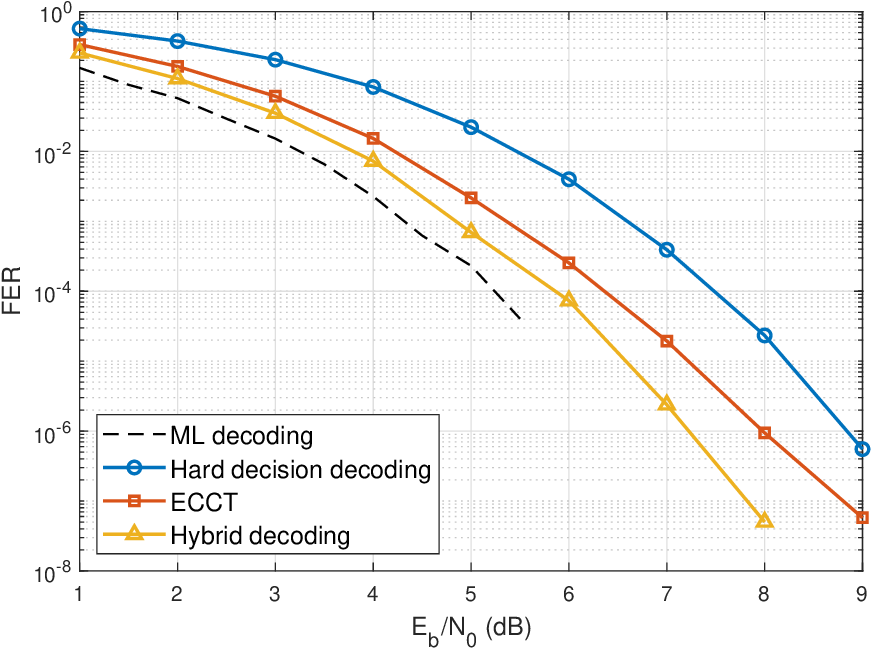}\label{fig:errorfloor_3116}}
        \hfil
        \subfloat[(63, 36) BCH code]{\includegraphics[width=0.4\textwidth]{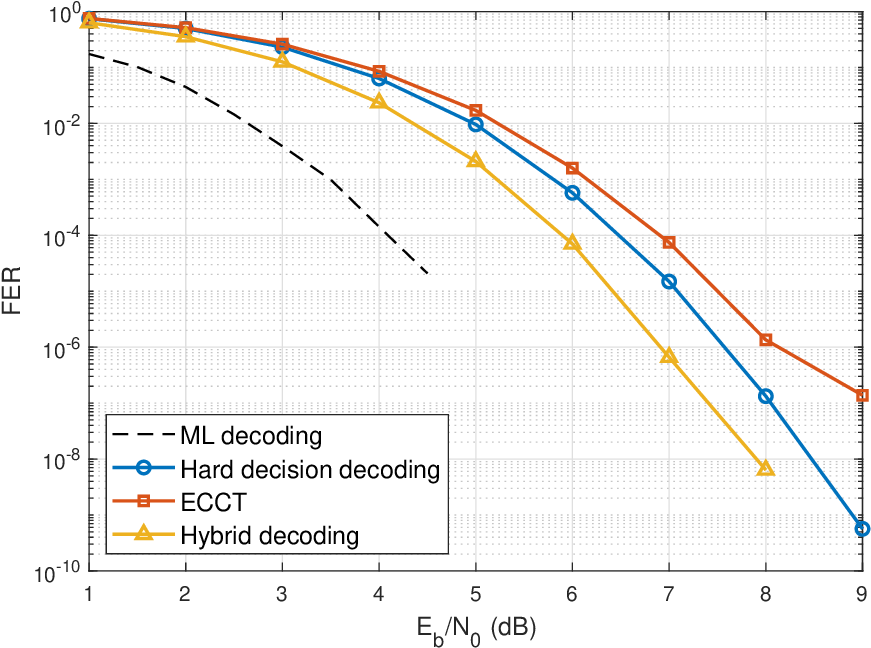}\label{fig:errorfloor_6336}}
        \hfil
        \subfloat[(63, 45) BCH code]{\includegraphics[width=0.4\textwidth]{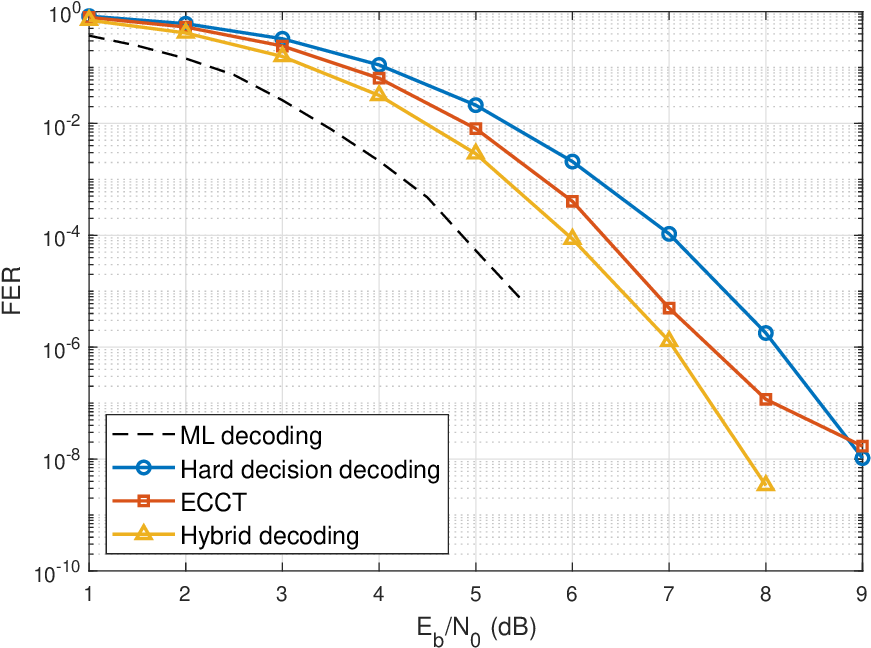}\label{fig:errorfloor_6345}}
        \hfil
        \subfloat[(63, 51) BCH code]{\includegraphics[width=0.4\textwidth]{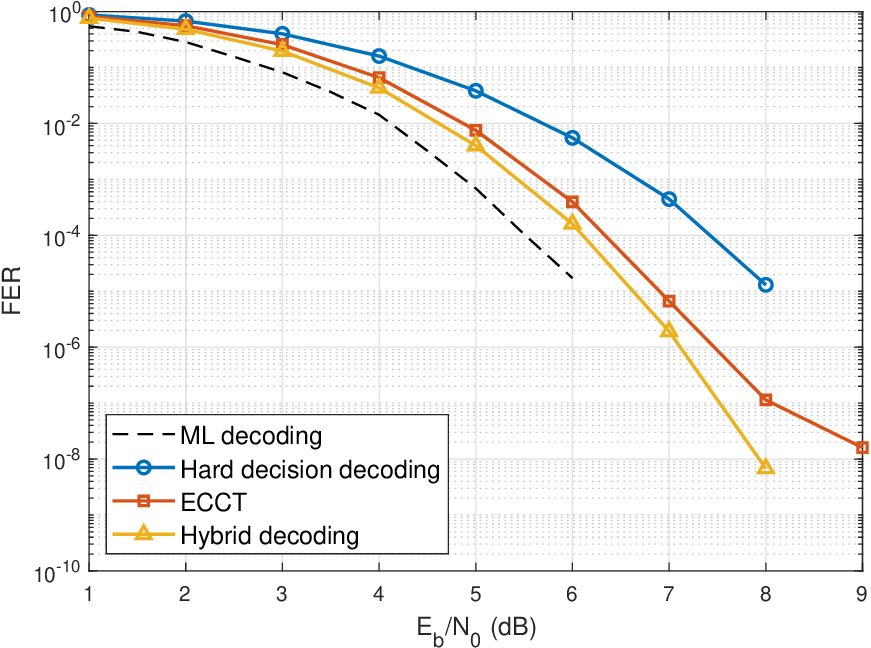}\label{fig:errorfloor_6351}}
        \hfil
        \caption{FER performance of ML decoding~\cite{Helmling2019database}, hard decision decoding, ECCT~\cite{Choukroun2022error}, and hybrid decoding.}
        \label{fig:hybrid_bch}
    \end{figure*} 
    
    In Lemma~\ref{lemma:loss_hard}, since the error probabilities $P_{\text{bu}}$ and $P_{\text{dd}}$ are non-trainable and fully determined by the hard decision decoder of underlying code, the ECCT-specific loss is as follows: 
    \begin{equation}\mathbb{E} \Big[
                u(d_H (\mathbf{X}, \widetilde{f}_{\boldsymbol{\theta}}(\mathbf{Y})) - t_c) d_H (\mathbf{X}, \widetilde{f}_{\boldsymbol{\theta}}(\mathbf{Y}) ) \Big|
                \mathbf{s}_{\text{pre}}\neq \mathbf{0} \Big].\label{eq:lemma1_equiv}
    \end{equation}
    In the full hybrid decoder loss~\eqref{eq:lemma1}, the constant term $P_{\text{bu}}$ stems from the use of the pre-decoder branch.
    Therefore, the decision to include a pre-decoder can be determined by the undetected error rate of the target code. 
    For short, high-rate codes, the probability of undetected errors is relatively high~\cite{Lin2004error}, making it preferable to adopt the hybrid architecture without a pre-decoder as shown in Fig.~\ref{fig:architecture}\subref{fig:architecture_post}.
    On the other hand, for codes with sufficiently long lengths and low rates, the benefit of using a pre-decoder--correction of all errors within the radius $t_c$--outweighs the additional constant term in the overall loss, making the architecture with pre-decoder as shown in Figs.~\ref{fig:architecture}\subref{fig:architecture_pre_post} and \subref{fig:architecture_pre} more favorable.
    
    In the loss~\eqref{eq:lemma1_equiv}, the pre-decoder is addressed through a conditional expectation, ensuring that ECCT is trained exclusively on received vectors $\mathbf{y}$ that the pre-decoder is unable to correct.
    In addition, the post-decoder is incorporated by multiplying the loss function with the step function $u(d_H(\mathbf{X},\widetilde{f}_{\boldsymbol{\theta}}(\mathbf Y))-t_c)$.
    ECCT trained with this loss aims to reduce errors to $t_c$ or fewer, rather than correcting all errors. 
    The remaining errors, which are less than or equal to $t_c$, are corrected by the post-decoder. 
    Furthermore, the loss~\eqref{eq:loss_hybrid2} can be upper bounded as in the following proposition.
    \begin{proposition}
    The loss~\eqref{eq:loss_hybrid2} can be upper bounded by
        \begin{align}
            \widetilde{\mathcal{L}}({\boldsymbol{\theta}})={}&\frac{P_{\text{dd}}}{\log{2}}\mathbb{E} \Big[ 
                u(d_H (\mathbf{X}, \widetilde{f}_{\boldsymbol{\theta}}(\mathbf{Y})) - t_c) \nonumber \\
                &   \times l_{\text{BCE}} (\mathbf{X}, f_{\boldsymbol{\theta}}(\mathbf{Y}) ) \Big|
                \mathbf{s}_{\text{pre}}\neq \mathbf{0}\Big]+P_{\text{bu}}.\label{eq:prop1}
        \end{align}
    \end{proposition}
    \begin{IEEEproof}
        Following Lemma~\ref{lemma:loss_hard}, the loss $\mathcal L(\boldsymbol{\theta})$ can be reformulated as~\eqref{eq:lemma1}.
        In~\eqref{eq:lemma1}, $d_H(\mathbf{x}, \widetilde{f}_{\boldsymbol{\theta}}(\mathbf{y}))=\sum_{i=1}^n d_H(x_i, \widetilde{f}_{\boldsymbol{\theta}}(\mathbf{y})_i)$ is equivalent to $\sum_{i=1}^n l_{0\text{--}1}((\tilde{z}_s)_i \cdot f_{\boldsymbol{\theta}}(\mathbf{y})_i)$, where $l_{0\text{--}1}(\alpha)$ is defined as $l_{0\text{--}1}(\alpha)=1$ if $\alpha\leq 0$ and $0$ otherwise.
        Here, $l_{0\text{--}1}(\alpha)$ is upper bounded by the logistic loss function as follows~\cite{Bartlett2006convexity}:
        \begin{equation}
        l_{0\text{--}1}((\tilde{z}_s)_i \cdot f_{\boldsymbol{\theta}}(\mathbf{y})_i) \le -\log_2(\sigma((\tilde{z}_s)_i \cdot f_{\boldsymbol{\theta}}(\mathbf{y})_i)).\nonumber
        \end{equation}        
        Since this logistic loss is equivalent to the scaled binary cross entropy loss in~\eqref{eq:bce}, the loss function \eqref{eq:lemma1} can be upper bounded by~\eqref{eq:prop1}.
    \end{IEEEproof}

    In practice, $\widetilde{\mathcal{L}}({\boldsymbol{\theta}}) $ can be computed using the following sample average, omitting the constant terms:
    \begin{align}\label{eq:hard_loss_practice}
    \widetilde{\mathcal{L}}({\boldsymbol{\theta}})
    &\simeq \frac{1}{M} \sum_{i=1}^M 
    \Big\{ u\big(d_H(\mathbf{x}^{(i)}, \widetilde{f}_{\boldsymbol{\theta}}(\mathbf{y}^{(i)})) - t_c\big) \nonumber \\
    &\quad \times l_{\text{BCE}}\big(\mathbf{x}^{(i)}, f_{\boldsymbol{\theta}}(\mathbf{y}^{(i)})\big) \Big\},
    \end{align}
    where $\mathbf{x}^{(i)}$ can be fixed as the zero codeword and $\mathbf{y}^{(i)}$ is sampled as $\mathbf{x}_s^{(i)} + \mathbf{z}^{(i)}$, retaining only the samples that satisfy $d_H(\mathbf{x}^{(i)},\tilde{\mathbf{y}}^{(i)})>t_c$. 
    Note that, to simplify the sampling process, we simply check whether $d_H(\mathbf x^{(i)}, \tilde{\mathbf y}^{(i)}) > t_c$, instead of checking the syndrome of the hard decision decoder by executing a hard decision decoder such as the Berlekamp-Massey algorithm~\cite{Berlekamp2015algebraic} for every sample.
    The noise variance $\sigma^2$ is uniformly sampled from the training SNR range.
    In addition, the step function $u(\alpha)$ is implemented via a straight-through estimator~(STE)~\cite{Bengio2013estimating} with the sigmoid function as a proxy, as described in~\cite{Xiao2020designing}. 
    The Hamming distance $d_H(\mathbf{x},\widetilde{f}_{\boldsymbol{\theta}}(\mathbf{y}))$ is estimated by multiplying the code length $n$ with the soft BER loss defined in~\cite{Lian2019learned}.
    In the case of the hybrid decoder architecture with only a pre-decoder, as shown in Fig.~\ref{fig:architecture}\subref{fig:architecture_pre}, the loss is computed by removing the step function from~\eqref{eq:hard_loss_practice}. For the hybrid decoder with only a post-decoder, as in Fig.~\ref{fig:architecture}\subref{fig:architecture_post}, the condition $d_H(\mathbf{x}^{(i)},\mathbf{y}^{(i)})>t_c$ is no longer checked.

    \begin{table*}[!t]
    \centering
    \caption{Comparison of FER of Decoders with Different Configurations in Error Floor Region}
    \begin{tabular}{lcccccc}
    \toprule
    \multicolumn{1}{l}{\multirow[c]{2}{*}[-1mm]{Methods}} & \multicolumn{3}{c}{(31, 16) BCH code} & \multicolumn{3}{c}{(63, 36) BCH code} \\
    \cmidrule(rl){2-4} \cmidrule(rl){5-7}
    {} & {\SI{6}{\decibel}} & {\SI{7}{\decibel}} & {\SI{8}{\decibel}} & {\SI{6}{\decibel}} & {\SI{7}{\decibel}} & {\SI{8}{\decibel}} \\
    \midrule
    ECCT~\cite{Choukroun2022error} & $2.54\mathrm{e}{-4}$ & $1.94\mathrm{e}{-5}$ & $9.42\mathrm{e}{-7}$ & $1.59\mathrm{e}{-3}$ & $7.44\mathrm{e}{-5}$ & $1.35\mathrm{e}{-6}$\\
    Pre + ECCT & $8.45\mathrm{e}{-4}$ & $9.23\mathrm{e}{-5}$ & $4.76\mathrm{e}{-6}$ & $2.53\mathrm{e}{-4}$ & $5.08\mathrm{e}{-6}$ & $3.26\mathrm{e}{-8}$\\
    ECCT + Post & $7.78\mathrm{e}{-5}$ & $4.25\mathrm{e}{-6}$ & $1.45\mathrm{e}{-7}$ & $1.42\mathrm{e}{-4}$ & $2.51\mathrm{e}{-6}$ & $1.76\mathrm{e}{-8}$\\
    Pre + ECCT + Post & $6.95\mathrm{e}{-4}$ & $7.59\mathrm{e}{-5}$ & $4.20\mathrm{e}{-6}$ & $7.69\mathrm{e}{-5}$ & $9.48\mathrm{e}{-7}$ & $8.29\mathrm{e}{-9}$\\
    ECCT + Post + loss~\eqref{eq:hard_loss_practice} & $\mathbf{7.32\mathrm{\mathbf{e}}{-5}}$ & $\mathbf{2.38\mathrm{\mathbf{e}}{-6}}$ & $\mathbf{5.02\mathrm{\mathbf{e}}{-8}}$ & -- & -- & --\\
    Pre + ECCT + Post + loss~\eqref{eq:hard_loss_practice} & -- & -- & -- & $\mathbf{7.00\mathrm{\mathbf{e}}{-5}}$ & $\mathbf{6.60\mathrm{\mathbf{e}}{-7}}$ & $\mathbf{6.38\mathrm{\mathbf{e}}{-9}}$\\
    \bottomrule
    \end{tabular}
    \label{tab:ablation}
    \end{table*}
    
    \section{Experimental Results}\label{sec:results}
    
    In this section, we present the experimental results for the hybrid decoding scheme, where the ECCT decoder is trained with the proposed loss function. 
    To evaluate its effectiveness, we measure the FER performance for four BCH codes: $(31, 16)$, $(63, 36)$, $(63, 45)$, and $(63, 51)$, which have error correcting capabilities of $t_c=3$, $t_c=5$, $t_c=3$, and $t_c=2$, respectively.
    In all simulations involving ECCT, the number of decoding layers $N$ is fixed to $6$, and the embedding dimension $d$ is set to $128$.
    Furthermore, the parity check matrices used in the experiments and maximum likelihood~(ML) decoding results are obtained from~\cite{Helmling2019database}. 
    
    
    Fig.~\ref{fig:hybrid_bch} shows the FER performance of the proposed hybrid decoder along with other decoders: ML decoding, hard decision decoding, and the original ECCT.
    The hybrid decoder combining both pre- and post-decoders is used for $(63, 36)$ BCH code, while the hybrid decoder with only the post-decoder is used for $(31, 16)$, $(63, 45)$, and $(63, 51)$ BCH codes, which yielded the best results.
    ECCTs in each hybrid decoder are trained using the proposed loss~\eqref{eq:hard_loss_practice}, with necessary adjustments based on the hybrid decoder architecture. 
    The results indicate that the hybrid decoder effectively mitigates the error floor, significantly outperforming the original ECCT in the high SNR region.
    Furthermore, even in the waterfall region, the hybrid decoder outperforms both the original ECCT and the hard decision decoder.
    By combining two distinct decoders and training them with the proposed loss, the hybrid decoding approach can achieve significant coding gains.

    To assess the impact of each component in the hybrid decoding framework, we compare the FER performance of different decoder configurations, as shown in Table~\ref{tab:ablation}. 
    In this table, ``Pre'' and ``Post'' represent hard decision pre- and post-decoders, respectively while ``ECCT'' refers to the original ECCT trained on the binary cross entropy loss~\eqref{eq:bce}.
    Among the various decoder architectures, the hybrid decoder with only a post-decoder yields the best performance for the $(31, 16)$ BCH code, whereas the architecture with both pre- and post-decoders performs best for the $(63, 36)$ BCH code.
    This result is attributed to the higher undetected error rate for short codes.
    Furthermore, for both codes, training the hybrid decoder with the proposed loss~\eqref{eq:hard_loss_practice} further improves FER performance compared to training with the binary cross entropy loss~\eqref{eq:bce}.

    \section{Conclusion}\label{sec:conclusion}
    
    In this paper, we presented the first observation of the error floor phenomenon in ECCT.
    To address this issue, we adopted hybrid decoding strategy that integrates ECCT with hard decision decoders, combined with the proposed loss function tailored to this architecture. 
    The hybrid decoding approach effectively broadens the range of correctable errors, thereby mitigating the error floor.
    In particular, the proposed loss function directs ECCT to focus on correcting critical errors, improving the effectiveness of the hybrid decoding process.
    Our simulation results support that the hybrid decoders trained with the proposed loss function can effectively lower the error floor, achieving improved FER performance in the high SNR region. 
    Additionally, the hybrid decoder significantly improves decoding performance even in the waterfall region.
    The hybrid decoding strategy, along with its training method that accounts for both pre- and post-decoders, can be generalized to other neural decoders. 
    This framework provides a flexible strategy for integrating neural decoders and conventional decoders to lower the error floor and improve overall decoding performance.
    In future work, exploring hybrid decoding strategies that combine ECCT with conventional BP decoders would be both valuable and promising.

    \section*{Acknowledgment}
     
    This work was partly supported by Institute of Information \& communications Technology Planning \& Evaluation (IITP) grant funded by the Korean government (MSIT) (RS-2024-00398449, Network Research Center: Advanced Channel Coding and Channel Estimation Technologies for Wireless Communication Evolution) and the National Research Foundation of Korea (NRF) grant funded by the Korean government (MSIT) (No. RS-2023-00212103).
    



    \appendices

    \bibliographystyle{IEEEtran}
    \bibliography{abrv,mybib}

\end{document}